\documentclass{iaus}
\usepackage{graphicx}

\newcommand{\Msun}{\mathrm{M}_{\odot}}
\def\apj{{\it ApJ}}
\def\aj{{\it AJ}}

\title[Globular Clusters in Hierarchical Cosmology]
{Dynamical Evolution of Globular Clusters \\ in Hierarchical Cosmology}

\author[Oleg Y. Gnedin \& Jos\'e L. Prieto]
{Oleg Y. Gnedin$^2$ \and Jos\'e L. Prieto$^2$}

\affiliation{$^1$University of Michigan, Department of Astronomy, 
                 Ann Arbor, MI 48109-1042, USA
             \\ {\tt ognedin@umich.edu}
             \\[\affilskip]
             $^2$The Ohio State University, Department of Astronomy, 
                 Columbus, OH 43210, USA
             \\ {\tt prieto@astronomy.ohio-state.edu}}

\pubyear{2008}
\volume{246}
\pagerange{1--5}
\jname{Dynamical Evolution of Dense Stellar Systems}
\editors{E. Vesperini, M. Giersz, A. Sills, eds.}
\begin{document}

\maketitle
\begin{abstract}
We probe the evolution of globular clusters that could form in giant
molecular clouds within high-redshift galaxies.  Numerical simulations
demonstrate that the large and dense enough gas clouds assemble
naturally in current hierarchical models of galaxy formation.  These
clouds are enriched with heavy elements from earlier stars and could
produce star clusters in a similar way to nearby molecular clouds.
The masses and sizes of the model clusters are in excellent agreement
with the observations of young massive clusters.  Do these model
clusters evolve into globular clusters that we see in our and external
galaxies?  In order to study their dynamical evolution, we calculate
the orbits of model clusters using the outputs of the cosmological
simulation of a Milky Way-sized galaxy.  We find that at present the
orbits are isotropic in the inner 50 kpc of the Galaxy and
preferentially radial at larger distances.  All clusters located
outside 10 kpc from the center formed in the now-disrupted satellite
galaxies.  The spatial distribution of model clusters is spheroidal,
with a power-law density profile consistent with observations.  The
combination of two-body scattering, tidal shocks, and stellar
evolution results in the evolution of the cluster mass function from
an initial power law to the observed log-normal distribution.
\end{abstract}

\firstsection
\section{Giant Molecular Clouds at High Redshift}

The outcomes of many proposed models of globular cluster formation
depend largely on the assumed initial conditions.  The collapse of the
first cosmological $10^6\ \Msun$ gas clouds, or the fragmentation of
cold clouds in hot galactic corona gas, or the agglomeration of
pressurized clouds in mergers of spiral galaxies could all, in
principle, produce globular clusters, but only if those conditions
realized in nature.  Similarly, while observational evidence strongly
suggests that all stars and star clusters form in molecular clouds,
the initial conditions for cloud fragmentation are a major uncertainty
of star formation models.

\begin{figure}
\centering
\includegraphics[width=3.1in]{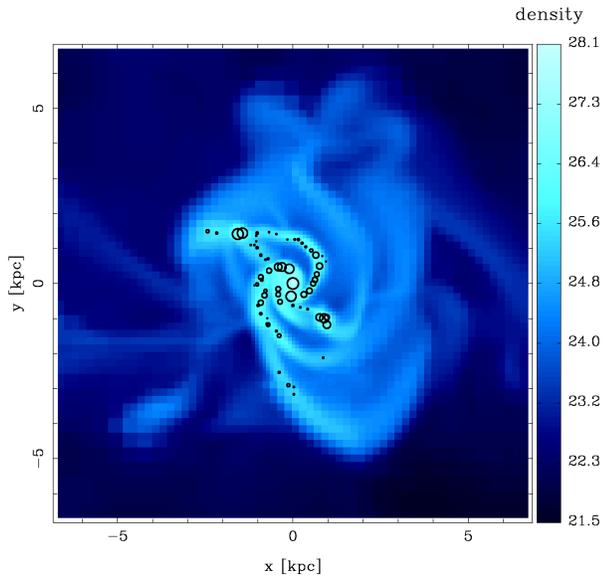}
\caption{A massive gaseous disk with prominent spiral arms, seen
  face-on at redshift $z=4$ in a process of active merging.  The gas
  density is projected over a $3.5$~kpc slice.  In our model star
  clusters form in giant gas clouds, shown by circles with the sizes
  corresponding to the cluster masses.  From Kravtsov \& Gnedin
  (2005).}
  \label{fig:gc}
\end{figure}

The only information that we actually have about the initial
conditions comes from the early universe, when primordial density
fluctuations set the seeds for structure formation.  These
fluctuations are probed directly by the anisotropies of the cosmic
microwave background radiation.  Cosmological numerical simulations
study the growth of these fluctuations via gravitational instability,
in order to understand the formation of galaxies and all other
structures in the Universe.  The simulations begin with tiny
deviations from the Hubble flow, whose amplitudes are set by the
measured power spectrum of the primordial fluctuations while the
phases are assigned randomly.  Therefore, each particular simulation
provides only a statistical description of a representative part of
the Universe, although current models successfully reproduce major
features of observed galaxies.

Kravtsov \& Gnedin (2005, \apj, 623, 650) attempted to construct a
first self-consistent model of star cluster formation, using an
ultrahigh-resolution gasdynamics cosmological simulation with the
Adaptive Refinement Tree code.  They identified supergiant molecular
clouds in high-redshift galaxies as the likely formation sites of
globular clusters.  These clouds assemble during gas-rich mergers of
progenitor galaxies, when the available gas forms a thin, cold,
self-gravitating disk.  The disk develops strong spiral arms, which
further fragment into separate molecular clouds located along the arms
as beads on a string (see Fig. \ref{fig:gc}).

In this model, clusters form in relatively massive galaxies, with the
total mass $M_{\rm host} > 10^{9}\ \Msun$, beginning at redshift $z
\approx 10$.  The mass and density of the molecular clouds increase
with cosmic time, but the rate of galaxy mergers declines steadily.
Therefore, the cluster formation efficiency peaks at a certain
extended epoch, around $z \approx 4$, when the Universe is only 1.5
Gyr old.  The host galaxies are massive enough for their molecular
clouds to be shielded from the extragalactic UV radiation, so that
globular cluster formation is unaffected by the reionization of cosmic
hydrogen.  As a result of the mass-metallicity correlation of
progenitor galaxies, clusters forming at the same epoch but in
different-mass progenitors have different metallicities, ranging
between $10^{-3}$ and $10^{-1}$ solar.  The mass function of model
clusters is consistent with a power law $dN/dM \propto M^{-\alpha}$,
where $\alpha = 2.0 \pm 0.1$, similar to the observations of nearby
young star clusters.

\begin{figure}
\centering
\includegraphics[height=2.25in]{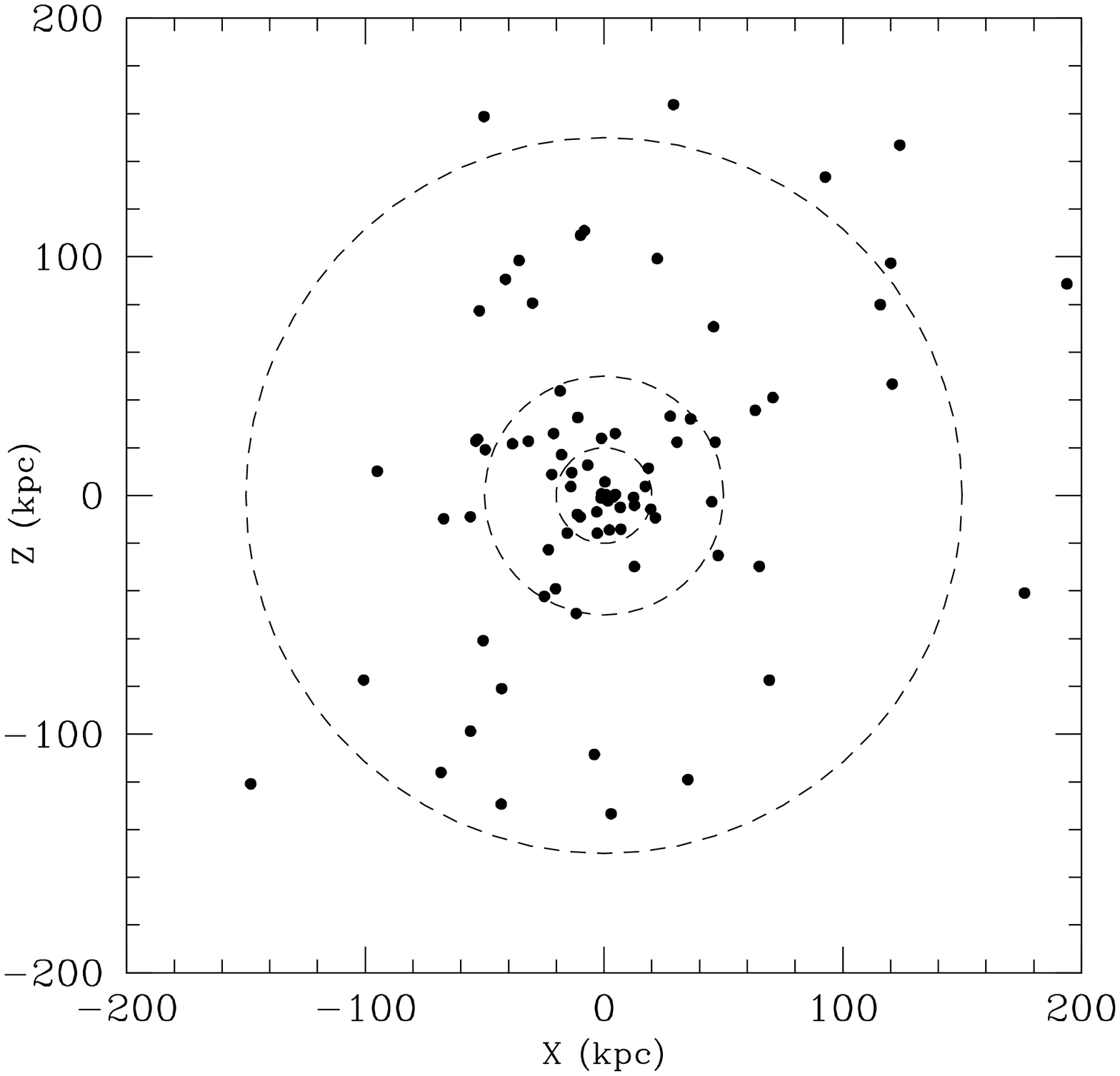}
\includegraphics[height=2.25in]{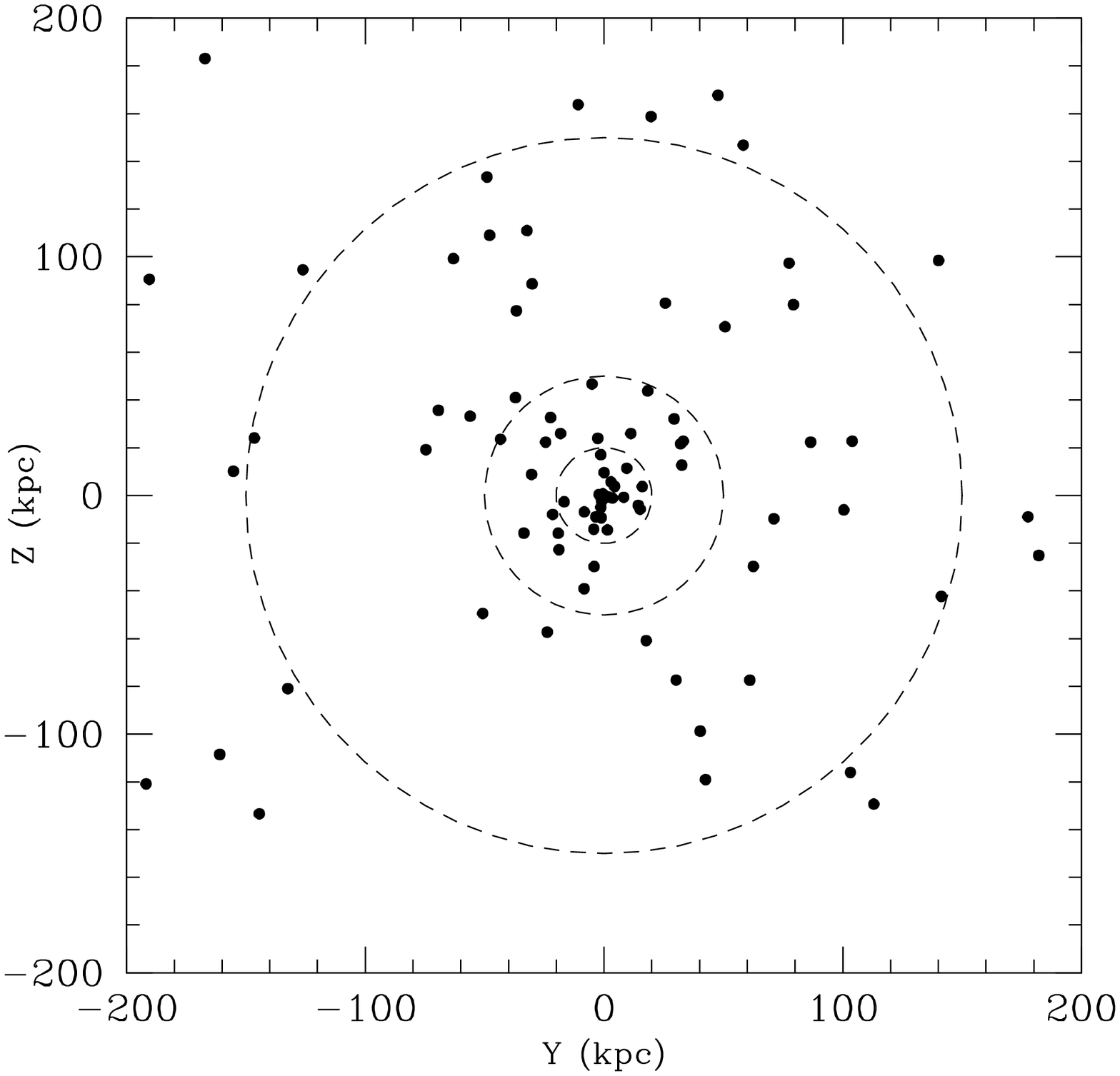}
\includegraphics[height=2.25in]{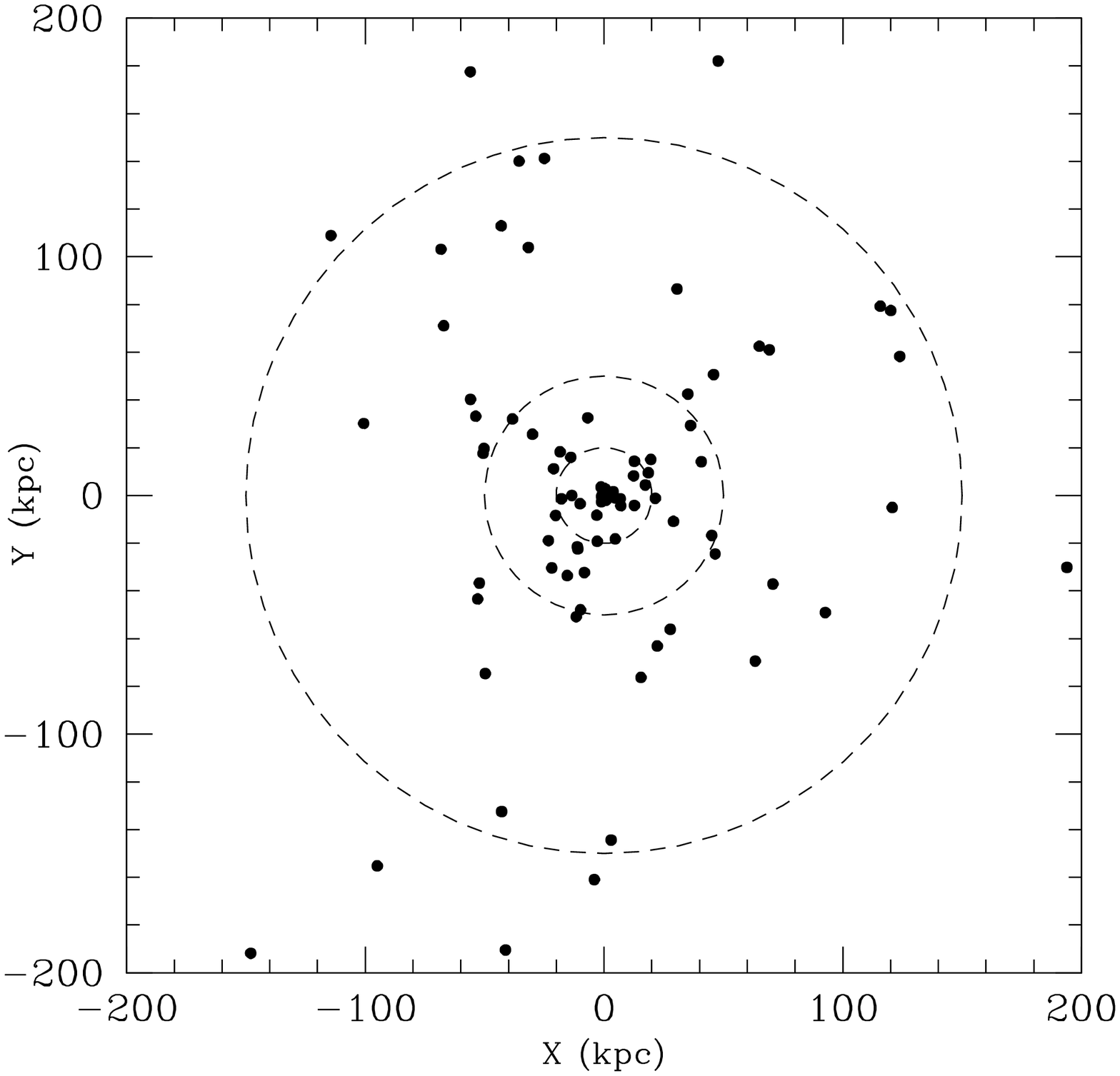}
\includegraphics[height=2.25in]{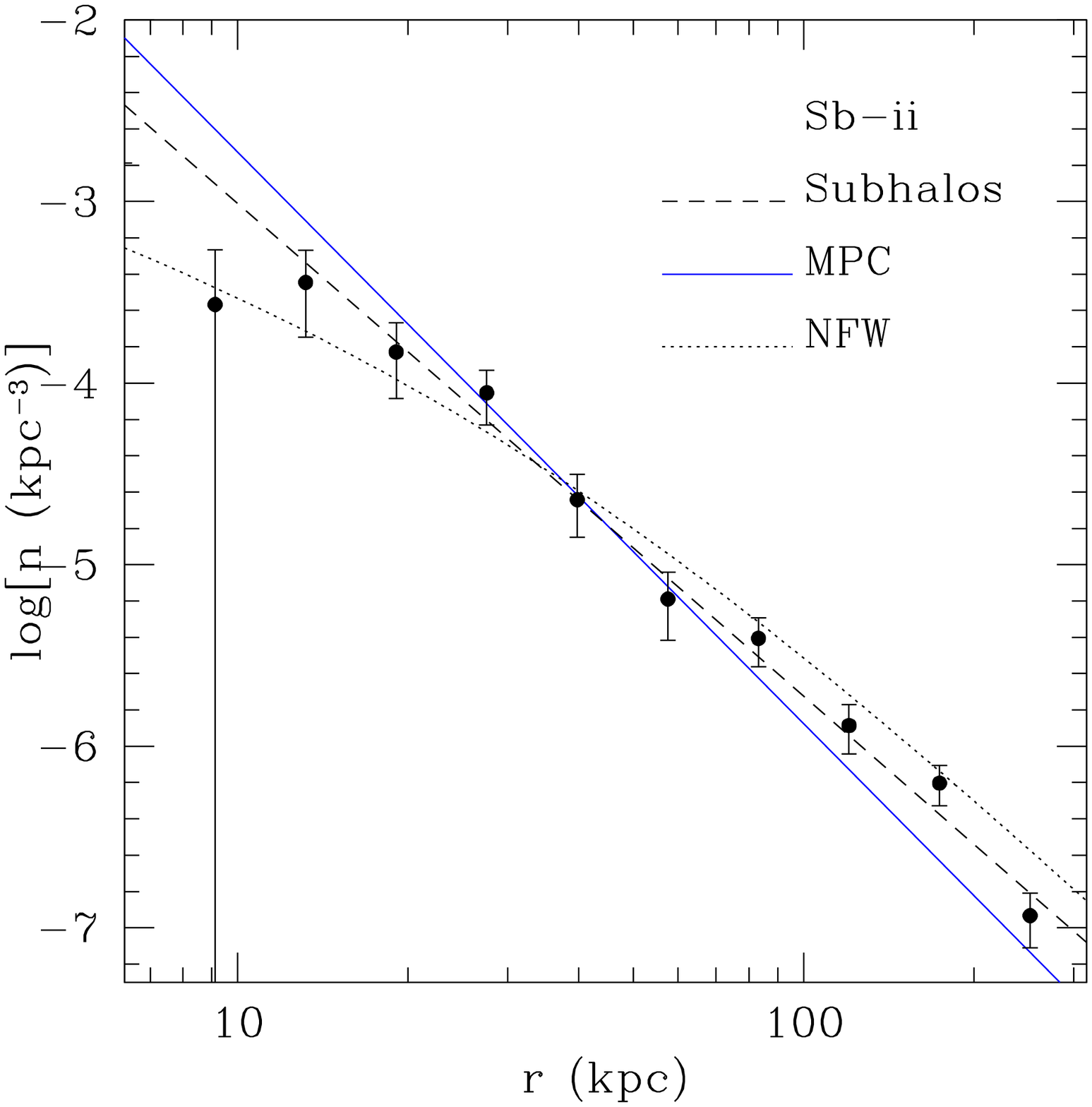}
\caption{Spatial distribution of surviving model clusters in the
  Galactic frame.  Dashed circles are to illustrate the projected
  radii of 20, 50, and 150 kpc.  The number density profile ({\it
  bottom right}) can be fit by a power law, $n(r)\propto r^{-2.7}$.
  The distribution of model clusters is similar to that of surviving
  satellite halos ({\it dashed line}) and smooth dark matter ({\it
  dotted line}).  It is also consistent with the observed distribution
  of metal-poor globular clusters in the Galaxy ({\it solid line}),
  plotted using data from the catalog of Harris (1996, \aj, 112,
  1487).}
  \label{fig:density}
\end{figure}

\begin{figure}
\centering 
\includegraphics[height=2.25in]{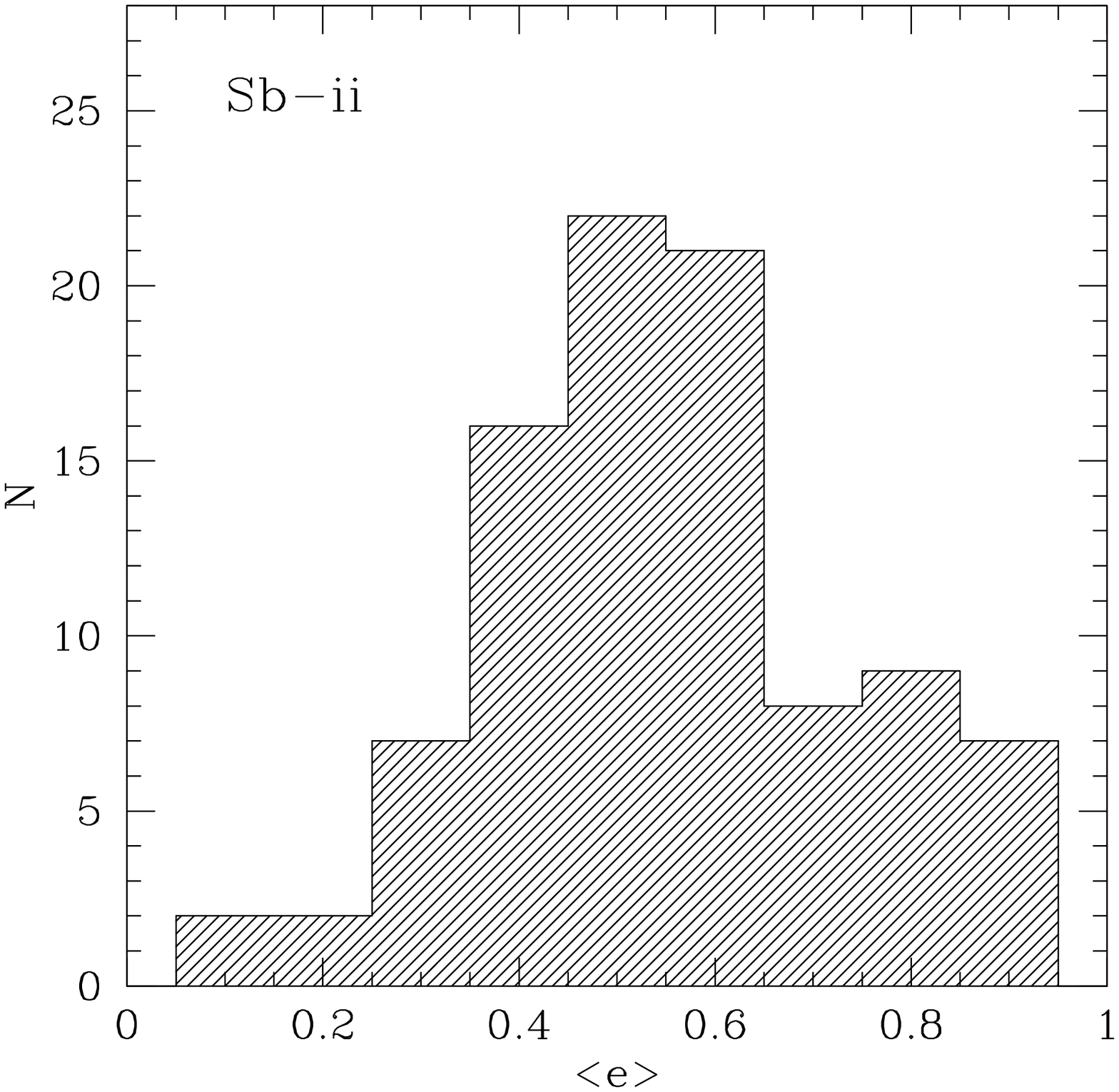}
\includegraphics[height=2.25in]{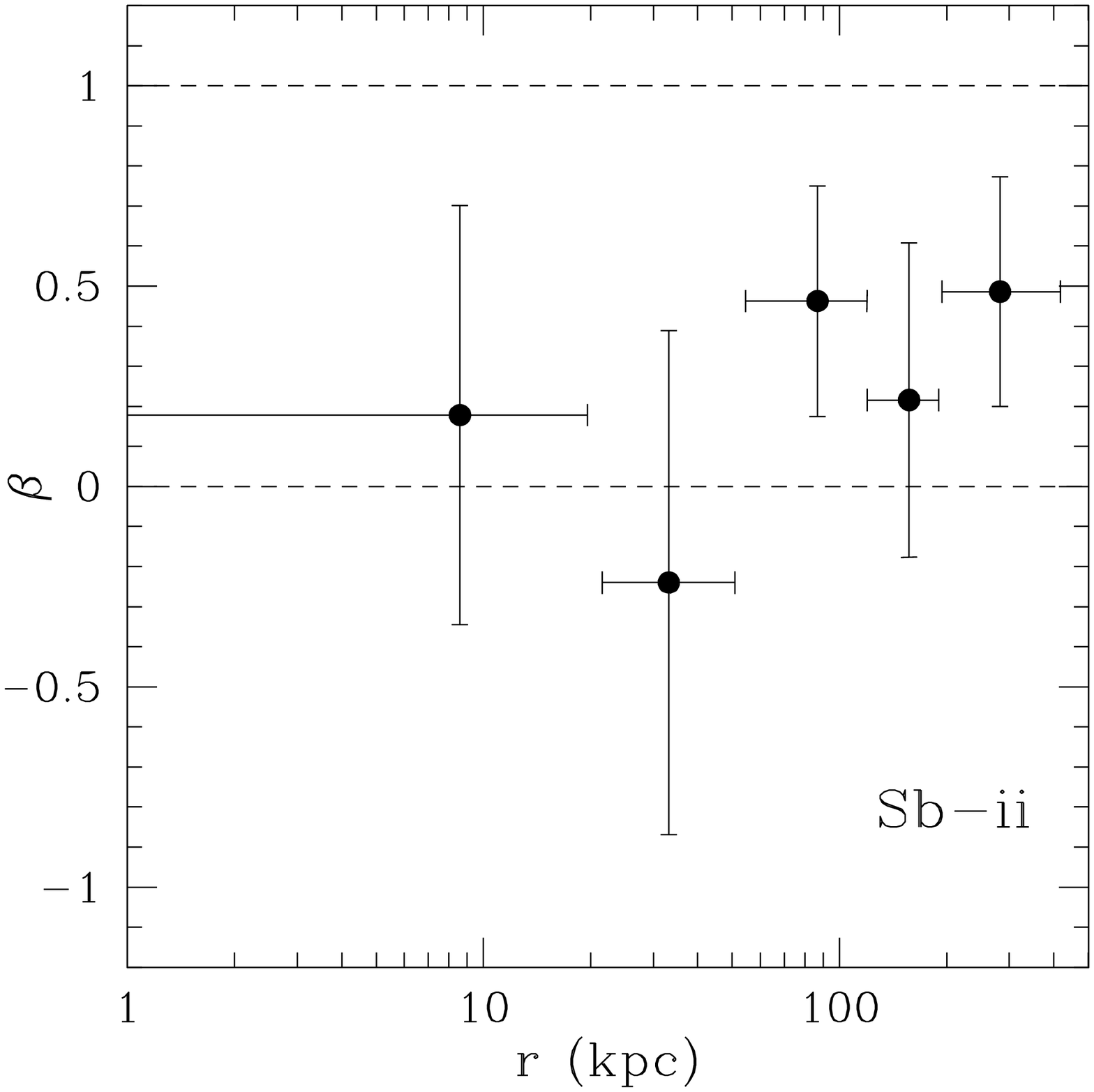}
\caption{{\it Left:} Average eccentricity distribution of the
  surviving model clusters.  {\it Right:} Anisotropy parameter
  $\beta$ as a function of radius.  Vertical errorbars represent the
  error of the mean for each radial bin, while horizontal errorbars
  show the range of the bin.  Horizontal dashed lines illustrate an
  isotropic ($\beta=0$) and a purely radial ($\beta=1$) orbital
  distributions.}
  \label{fig:ecc}
\end{figure}

\section{Orbits of Globular Clusters}

We adopt this model to set up the initial positions, velocities, and
masses for our globular clusters.  We then calculate cluster orbits
using a separate collisionless $N$-body simulation described in
Kravtsov, Gnedin \& Klypin (2004, \apj, 609, 482).  This is necessary
because the original gasdynamics simulation was stopped at $z \approx
3.3$, due to limited computational resources.  By using the $N$-body
simulation of a similar galactic system, but complete to $z=0$, we are
able to follow the full dynamical evolution of globular clusters until
the present epoch.  We use the evolving properties of all progenitor
halos, from the outputs with a time resolution of $\sim 10^8$ yr, to
derive the gravitational potential in the whole computational volume
at all epochs.  We convert a fraction of the dark matter mass into
flattened disks, in order to model the effect of baryon cooling and
star formation on the galactic potential.  We calculate the orbits of
globular clusters in this potential from the time when their host
galaxies accrete onto the main (most massive) galaxy.  Using these
orbits, we calculate the dynamical evolution of model clusters,
including the effects of stellar mass loss, two-body relaxation, tidal
truncation, and tidal shocks.

We consider several possible scenarios, some with all clusters forming
in a short interval of time around redshift $z=4$, and others with a
continuous formation of clusters between $z=9$ and $z=3$.  Below we
discuss the spatial and kinematic distributions of globular clusters
in the best-fit model with the synchronous formation at $z=4$.

In our model, all clusters form on nearly circular orbits within the
disks of progenitor host galaxies.  Depending on the subsequent
trajectories of the hosts, clusters form three main subsystems at
present time.  {\it Disk clusters} formed in the most massive
progenitor that eventually hosts the present Galactic disk.  These
clusters, found within the inner 10 kpc, do not actually stay on
circular orbits but instead are scattered to eccentric orbits by
perturbations from accreted galactic satellites.  {\it Inner halo
clusters}, found between 10 and 60 kpc, came from the now-disrupted
satellite galaxies.  Their orbits are inclined with respect to the
Galactic disk and are fairly isotropic.  {\it Outer halo clusters},
beyond 60 kpc from the center, are either still associated with the
surviving satellite galaxies, or were scattered away from their hosts
during close encounters with other satellites and consequently appear
isolated.

Mergers of progenitor galaxies ensure the present spheroidal
distribution of the globular cluster system (Fig. \ref{fig:density}).
Most clusters are now within 50 kpc from the center, but some are
located as far as 200 kpc.  The azimuthally-averaged space density of
globular clusters is consistent with a power law, $n(r)\propto
r^{-\gamma}$, with the slope $\gamma \approx 2.7$.  Since all of the
distant clusters originate in progenitor galaxies and share similar
orbits with their hosts, the distribution of the clusters is almost
identical to that of the surviving satellite halos.  This power law is
similar to the observed distribution of the metal-poor ($\mbox{[Fe/H]}
< -0.8$) globular clusters in the Galaxy.  Such comparison is
appropriate, for our model of cluster formation at high redshift
currently includes only low metallicity clusters ($\mbox{[Fe/H]} \le
-1$).  Thus the formation of globular clusters in progenitor galaxies
with subsequent merging is fully consistent with the observed spatial
distribution of the Galactic metal-poor globulars.

Figure \ref{fig:ecc} shows the kinematics of model clusters.  Most
orbits have moderate average eccentricity, $0.4 < \left< e \right> <
0.7$, expected for an isotropic distribution.  The anisotropy
parameter, $\beta = 1 - v_t^2/2v_r^2$, is indeed close to zero in the
inner 50 kpc from the Galactic center.  At larger distances, cluster
orbits tend to be more radial.  There, in the outer halo, host
galaxies have had only a few passages through the Galaxy or even fall
in for the first time.

\begin{figure}
\centering \includegraphics[height=2.65in]{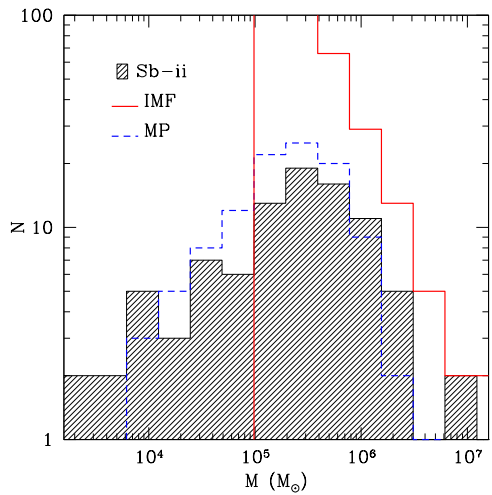}
\caption{Evolution of the mass function of clusters in our best-fit
  model from an initial power law ({\it solid line}) to a peaked
  distribution at present ({\it histogram}), including mass loss due
  to stellar evolution, two-body relaxation, and tidal shocks.  For
  comparison, dashed histogram shows the mass function of metal-poor
  globular clusters in the Galaxy.}
  \label{fig:mf}
\end{figure}

\section{Evolution of the Globular Cluster Mass Function}

Using these orbits, we now calculate the cluster disruption rates.
Sophisticated models of the dynamical evolution have been developed
using $N$-body simulations as well as orbit-averaged Fokker-Planck and
Monte Carlo models.  Several processes combine and reinforce each
other in removing stars from globular clusters: stellar mass loss,
two-body scattering, external tidal shocks, and dynamical friction of
cluster orbits.  The last three are sensitive to the external tidal
field and therefore, to cluster orbits.  While a general framework for
all these processes has been worked out in the literature, the
knowledge of realistic cluster orbits is essential for accurate
calculations of the disruption.

Figure \ref{fig:mf} shows the transformation of the cluster mass
function from an initial power law, $dN/dM \propto M^{-2}$, into a
final bell-shaped distribution.  In this model all globular clusters
form at the same redshift, $z=4$, or about 12 Gyr ago.  The half-mass
radii, $R_h$, are set by the condition that the median density,
$M/R_h^3$, is initially the same for all clusters and remains constant
as a function of time.  Over the course of their evolution, numerous
low-mass clusters are disrupted by two-body relaxation while the
high-mass clusters are truncated by tidal shocks.  The present mass
function is in excellent agreement with the observed mass function of
the Galactic metal-poor clusters.

This result by itself is not new.  Previous studies of the evolution
of the cluster mass function have found that almost any initial
function can be turned into a peaked distribution by the combination
of two-body relaxation and tidal shocks.  However, the efficiency of
these processes depends on the cluster mass and size, $M(t)$ and
$R_h(t)$.  {\it The new result is that we find that not all initial
relations $R_h(0) - M(0)$ and not all evolutionary scenarios $R_h(t) -
M(t)$ are consistent with the observed mass function.}

Consider two examples.  (i) If the half-mass radius $R_h$ is kept
fixed for clusters of all masses and at all times, the median density
$M(t)/R_h^3$ decreases as the clusters lose mass.  Two-body scattering
becomes less efficient and spares many low-mass clusters, while tidal
shocks become more efficient and disrupt most high-mass clusters.  The
final distribution is severely skewed towards small clusters.  (ii) If
the size is assumed to evolve in proportion to the mass, $R_h(t)
\propto M(t)$, the cluster density increases with time.  As a result,
all of the low-mass clusters are disrupted by the enhanced two-body
relaxation, while the high-mass clusters are unaffected by the
weakened tidal shocks.  The final distribution is skewed towards
massive clusters.  

Only our best-fit model with $M(t)/R_h^3(t) = const$ successfully
reproduces the mass function and spatial distribution of metal-poor
globular clusters in Galaxy.  We are now investigating the formation
of metal-rich clusters in galactic mergers at lower redshifts.

\end{document}